\documentclass[epj,referee]{svjour}
\usepackage{graphics}
\begin{document}

\title{Promoting cooperation in social dilemmas via simple coevolutionary rules}

\author{Attila Szolnoki\inst{1} and Matja{\v z} Perc\inst{2}}

\institute{Research Institute for Technical Physics and Materials Science, P.O. Box 49, H-1525 Budapest, Hungary
\and Department of Physics, Faculty of Natural Sciences and Mathematics, University of Maribor, Koro{\v s}ka cesta 160, SI-2000 Maribor, Slovenia}

\date{Received: date / Revised version: date}

\authorrunning{A. Szolnoki and M. Perc}
\titlerunning{Promoting cooperation in social dilemmas via simple coevolutionary rules}

\abstract{We study the evolution of cooperation in structured populations within popular models of social dilemmas, whereby simple coevolutionary rules are introduced that may enhance players abilities to enforce their strategy on the opponent. Coevolution thus here refers to an evolutionary process affecting the teaching activity of players that accompanies the evolution of their strategies. Particularly, we increase the teaching activity of a player after it has successfully reproduced, yet we do so depending on the disseminated strategy. We separately consider coevolution affecting either only the cooperators or only the defectors, and show that both options promote cooperation irrespective of the applied game. Opposite to intuitive reasoning, however, we reveal that the coevolutionary promotion of players spreading defection is, in the long run, more beneficial for cooperation than the likewise promotion of cooperators. We explain the contradictory impact of the two considered coevolutionary rules by examining the differences between resulting heterogeneities that segregate participating players, and furthermore, demonstrate that the influential individuals completely determine the final outcome of the games. Our findings are immune to changes defining the type of considered social dilemmas and highlight that the heterogeneity of players, resulting in a positive feedback mechanism, is a fundamental property promoting cooperation in groups of selfish individuals.
}

\maketitle

\section{Introduction}\label{sec:intro}
Social dilemmas constitute situations in which private or personal interests are at odds with the collective wellbeing \cite{macy02,skyrms}. Indeed, such situations are commonplace in the real world. While selfish individuals mostly champion their own prosperity and success, communities hosting them eventually require attention in form of altruistic acts as well. Failing to acknowledge this seemingly very reasonable demand begets havoc and leads to distress. Such tensions building up due to discrepancies between personal comfort and social welfare are at the core of all social dilemmas. Mutually cooperative behavior \cite{axel84} is considered an universal escape hatch out of the crux, whereby prosperity of individuals is partially sacrificed and put second place on behalf of common interests. Evolutionary game theory \cite{ms82,hofbauer88,gintis00,nowak06} provides competent mathematical tools to address and study different social dilemmas, and has since its establishment advanced to the preferred way of approaching the problem across many areas of social and natural sciences. Although the prisoner's dilemma still seems to be ahead of other games in terms of research efforts devoted to them, alternatives are catching up as the actual payoff ranking was dubbed difficult \cite{milinski97,turner99} and the established results related to the promotion of cooperation by spatial structure \cite{nowak92,lindgren94,schweitzer02} have been questioned within the related snowdrift game \cite{hauert04}. Indeed, several recent works have focused on different social dilemmas in order to extend the scope of presented findings and reiterate their importance \cite{santos05,santos06,pach06,szolnoki08,wang_pre08,fort_epl08}. Given that seemingly minute difference between payoff rankings can have a rather profound effect on the success of participating strategies, this trend is desirable and should be continued while aspiring for cooperation within the evolutionary game theory.

Research published in recent years has made it clear that heterogeneities amongst players might play a crucial role in the evolution of cooperation. For example, scale-free networks have been recognized as potent promoters of cooperative behavior in the prisoner's dilemma, snowdrift and the stag-hunt game \cite{santos05,santos06}. In fact, evolutionary games on complex networks \cite{abramson01,ebel02,wu05,tomassini05,szabo06,poncela07,wang07,masuda07,rong07,chen08,sz08,vukov08} in general tend to promote cooperation past the boundaries imposed by regular lattices, as comprehensively reviewed in \cite{szabopr06}. Similarly, heterogeneities in strategy adoption probabilities can also enhance cooperation \cite{kim02,wu06,chen_axv07,guan_pre07}, especially if the strategy adoption is favored from the more influential players \cite{szolnoki07}. Heterogeneities can also be introduced directly to payoffs in terms of dynamical \cite{perc06,perc07,tanimoto07} or quenched \cite{perc08} noise, whereby cooperators are promoted as well provided the uncertainties are adequately adjusted and distributed. While virtually all above approaches can be interpreted as justified within one or another real life scenario, the question remaining is how can we avoid introducing the heterogeneities manually and allow them to emerge spontaneously as an accompanying part of the evolutionary process affecting the distribution of strategies. Similar questions have been addressed by Pacheco {\it et. al.} introducing dynamical linking in games on graphs \cite{pach06,pachecojtb06}, as well as by Poncela {\it et. al.} elaborating on the emergence of complex networks via evolutionary preferential attachment \cite{poncela08}, and by Pestelacci {\it et. al.} who studied the evolution of cooperation and coordination \cite{pestelacci_axv08}. Notably, somewhat earlier studies already employed random or intentional rewiring procedures \cite{zimmermann04,eguiluz_ajs,perc06b} to elaborate on the sustainability of cooperation within social dilemmas. Besides trying to generate the desired heterogeneities via evolving interaction networks, an alternative coevolutionary approach affecting the diversity of players teaching activities has recently been proposed \cite{szolnoki08}. The teaching activity of each individual quantifies its ability to enforce its strategy on the opponent, whereby in accordance with logical reasoning, active individuals are more likely to reproduce than players characterized with a low teaching activity. Resulting heterogeneities from the coevolutionary process were found to promote cooperation in the prisoner's dilemma and the snowdrift game. Despite of the obvious differences between the coevolution of networks and teaching activity, however, we argue that both have a similar impact on the evolution of cooperation in that an increase of the teaching activity and an increase of degree both indirectly make the player more influential within the population. Hence, results presented by Poncela et al. \cite{poncela08}, for example, are related to our work \cite{szolnoki08} in that they both incorporate a rich-gets-riches scheme despite of the fact that the former approach involves growth as well.

Nevertheless, it is not straightforward to assume that all donors should be promoted by increasing their teaching activities after a successful strategy transfer. In particular, one may argue that the act of promotion should depend on the type of the transferred strategy. This distinction is motivated by the essential conflict between personal welfare and common good underpinning all social dilemmas, and as we will show below, may indeed vitally affect the evolutionary success of participating strategies. To clarify this issue, we here study two different coevolutionary rules separately. In both cases the game starts by assigning the same low teaching activity to all players, and subsequently, parallel with the evolution of strategies the teaching activity is evolved as well. The difference is that in one case this coevolutionary rule applies only to cooperators and in the other it applies only to defectors.
While both coevolutionary rules do promote cooperative behavior, the coevolutionary promotion of players spreading defection is, in the long run, more beneficial for cooperation than the likewise direct promotion of cooperators. Although being rather surprising, we shed light on this result by examining the differences between resulting heterogeneities in teaching activity that segregate the players into groups of active and virtually inactive individuals. Moreover, we reveal that the active individuals fully dominate the strategy transfers and thus completely determine the final outcome of the games in terms of average strategy densities on the grid. Given the simplicity of the considered coevolutionary rules, this work outlines a new and transparent route towards cooperation in structured populations of self-interested individuals, and it reiterates the importance and potency of influential players in maintaining a high level of overall welfare irrespective of the type of the governing social dilemma.

The remainder of this paper continuous as follows. In the next section we describe the three studied social dilemmas, as well as the employed protocols for the coevolution of teaching activity and details of performed calculations. Section~\ref{sec:res} features the results, whereas in the last section we summarize them and briefly discuss their implications.

\section{Social dilemmas and coevolutionary rules}\label{sec:mod}
Social dilemmas considered within this study are the spatial prisoner's dilemma, the spatial snowdrift and the spatial stag-hunt game. In all three games players can choose either to cooperate or defect, whereby mutual cooperation yields the reward $R$, mutual defection leads to punishment $P$, and the mixed choice gives the cooperator the sucker's payoff $S$ and the defector the temptation $T$. Depending on the rank of these four payoffs we have the prisoner's dilemma game if $T>R>P>S$, the snowdrift game if $T>R>S>P$, and the stag-hunt game if $R>T>P>S$. For simplicity, we here take $R=1$ and $P=0$, which imposes boundaries on the remaining two payoffs of the form $-1 \leq S \leq 1$ and $0 \leq T \leq 2$ \cite{santos06}. The rank of the four payoffs and the latter boundaries uniquely determine intervals of $S$ and $T$ for each game. To eschew additional effects of complex network topologies, and thus focus solely on the impact of introduced coevolutionary rules, we employ a regular $L \times L$ square lattice with periodic boundary conditions irrespective of which social dilemma applies. Initially, a player on the site $x$ is designated as a cooperator ($s_x=C$) or defector ($D$) with equal probability, and the game is iterated in accordance with the Monte Carlo simulation procedure comprising the following elementary steps. First, a randomly selected player $x$ acquires its payoff $p_x$ by playing the game with its four nearest neighbors. Next, one randomly chosen neighbor, denoted by $y$, also acquires its payoff $p_y$ by playing the game with its four neighbors. Last, player $x$ tries to enforce its strategy $s_x$ on player $y$ in accordance with the probability
\begin{equation}
W(s_y \rightarrow s_x)=w_x \frac{1}{1+\exp[(p_y-p_x)/K]}\,\,,
\label{eq:prob}
\end{equation}
where $K$ denotes the amplitude of noise \cite{toke98} or its inverse $(1/K)$ the so-called intensity of selection \cite{traulsen07,traulsen08}, and $w_x$ characterizes the level of teaching activity of player $x$ \cite{szolnoki07}. One full Monte Carlo generation involves all players having a chance to pass their strategies to the neighbors once on average. At the same time $w_x$ is also subjected to an evolutionary process in accordance with the following protocol that applies to all three social dilemmas alike. Initially, all players are given the minimal influence factor $w_x = w_m \ll 1$, thus assuring a nonpreferential setup of the game. Note, however, that $w_m$ must be positive in order to avoid frozen states, and hence we use $w_m = 0.01$ throughout this study. Next, every time player $x$ succeeds in enforcing its strategy on $y$ the influence $w_x$ may be increased by a constant positive value $\Delta w \ll 1$ according to $w_x \rightarrow w_x + \Delta w$. Presently we use $\Delta w = 0.1$, which gives a good compromise between fine increment and noticeable promotion in case of a successful strategy transfer. Finally, the evolution of influence is stopped as soon as one $w_x$ reaches $1$. As highlighted above, the described rule for the evolution of teaching activity is implemented depending on the strategy of player $x$ at the time of a successful reproduction. Namely, we separately study  the case where the coevolutionary rule applies only if $s_x=C$ or only if $s_x=D$. Thus, we consider coevolutionary promotion affecting either only the cooperators (rule A) or only the defectors (rule B). In both cases the final distribution of $w_x$ is obtained within a short period of time [typically around $100$ to $1000$ full Monte Carlo generations]. In spite of its simplicity the proposed protocol for the coevolution of teaching activity is remarkably robust, delivering conclusive results with respect to the promotion of cooperation.

We have verified the validity of this simple coevolutionary rule by employing several alternative, albeit slightly more sophisticated, coevolutionary protocols. For example, we have considered the case by which $w_x$ was allowed to grow also past $1$, only that then $w_x$ was normalized according to $w_x \rightarrow \frac{w_x}{w_{max}}$ ($w_{max}>1$ being the maximal out of all $w_x$ at any given instance of the game) to assure that the teaching activity remained bounded to the unit interval. Another alternative was not to use a fix value of $\Delta w$, but one that varies in time so that the $w_x=1$ limit is never reached. In particular, $\Delta w = (1-w_{max})/N$, where $N$ is a constant, can be used to achieve this. The latter coevolutionary rule may mimic the possibility that the award decreases in time, or in other words, that pioneers benefit substantially more from their work than their followers (see \cite{newman_ax} for an interesting recent study of this effect). However, no matter the details, these alternatives do not yield significantly different results from the here employed simplest version, but deviate merely in technical details, such as the required relaxation times, or relative differences in the final densities of cooperators on the spatial grid. It is also worth noting that experimenting with different coevolutionary rules may artificially promote cooperative behavior where in fact the environment alone would be insufficient to sustain cooperation. A model where such effects could be observed is given when, besides promoting the successful player by increasing its teaching activity, the unsuccessful player (if it fails to reproduce when attempted) is downgraded via $w_x \rightarrow w_x - \Delta w$.
In such a case an extremely high cooperation level can be obtained by direct promotion of cooperators, which is an artificial effect. Furthermore, it is possible to consider a natural decay of teaching activity that applies equally to all players, but such a process is likely to result in a relatively homogeneous distribution of $w$, which is not particularly beneficial for cooperation if compared to the originally proposed model. We would like to emphasize that our goal was to present the essence of the considered coevolutionary rules, and not so much burden with the actual modeling of potential real life scenarios. Thus, we kept the model as simple as possible, thereby allowing an efficient examination and strip it from potential artificial influences that otherwise might have gone unnoticed.

Monte Carlo results presented below were obtained on populations comprising $400 \times 400$ to $1000 \times 1000$ individuals, whereby the stationary fraction of cooperators $\rho_C$ was determined within $5\cdot10^5$ to $3\cdot10^6$ full Monte Carlo generations after sufficiently long transients were discarded. Moreover, due to the much shorter temporal scale characterizing the evolution of teaching activity and its resulting heterogeneous distribution, final results were additionally averaged over $10$ to $50$ independent runs for each set of parameter values in order to assure accuracy.

\section{Results}\label{sec:res}

\begin{figure}
\resizebox{0.95\columnwidth}{!}{\includegraphics{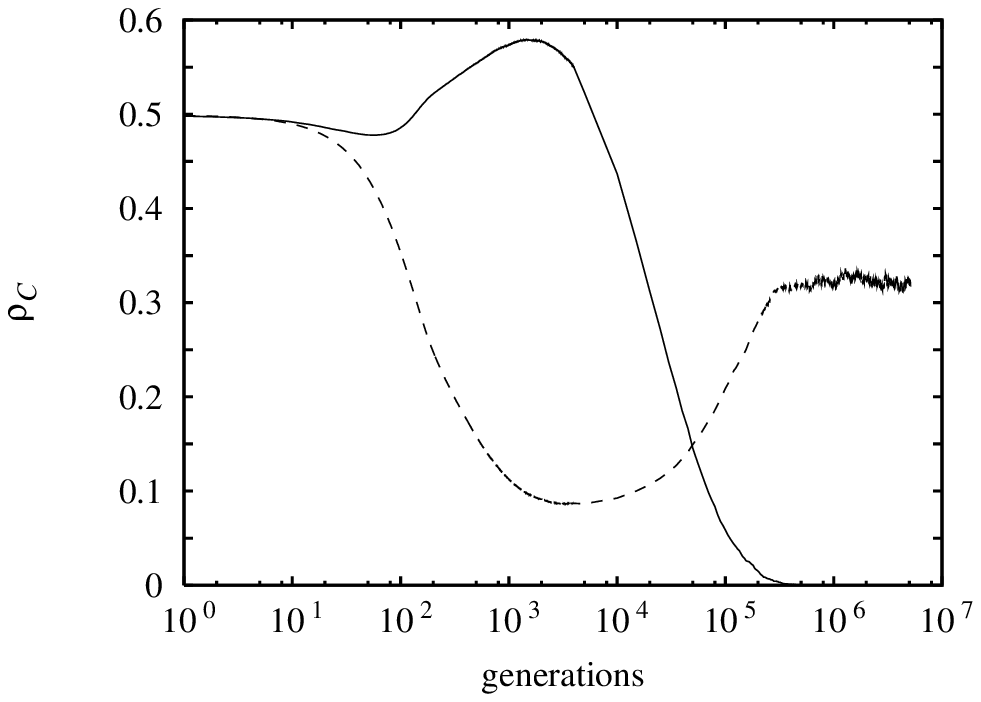}}
\resizebox{0.95\columnwidth}{!}{\includegraphics{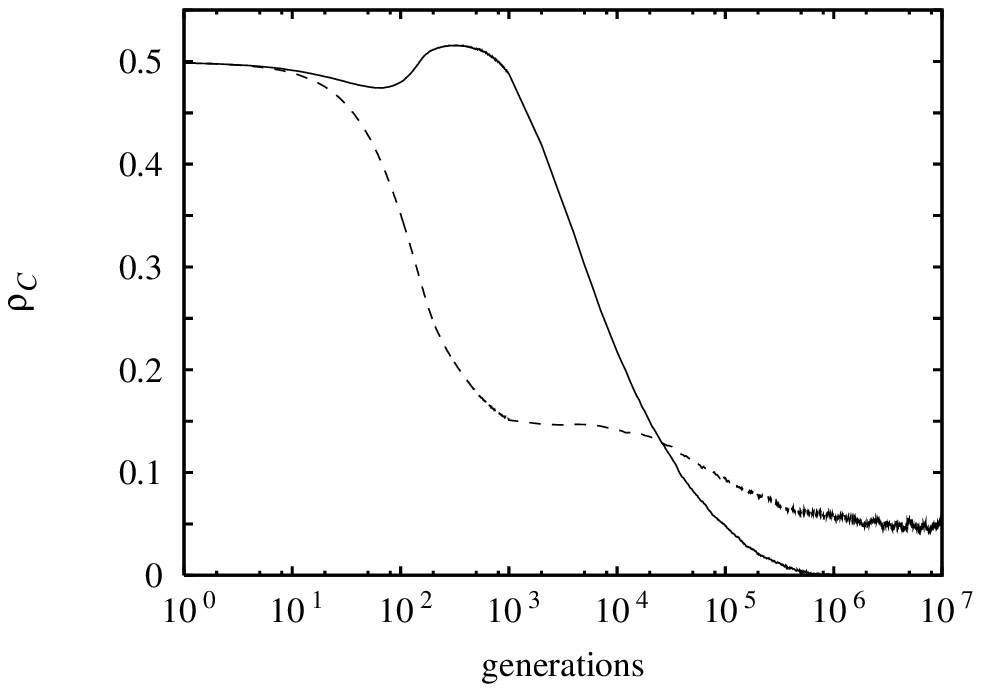}}
\resizebox{0.95\columnwidth}{!}{\includegraphics{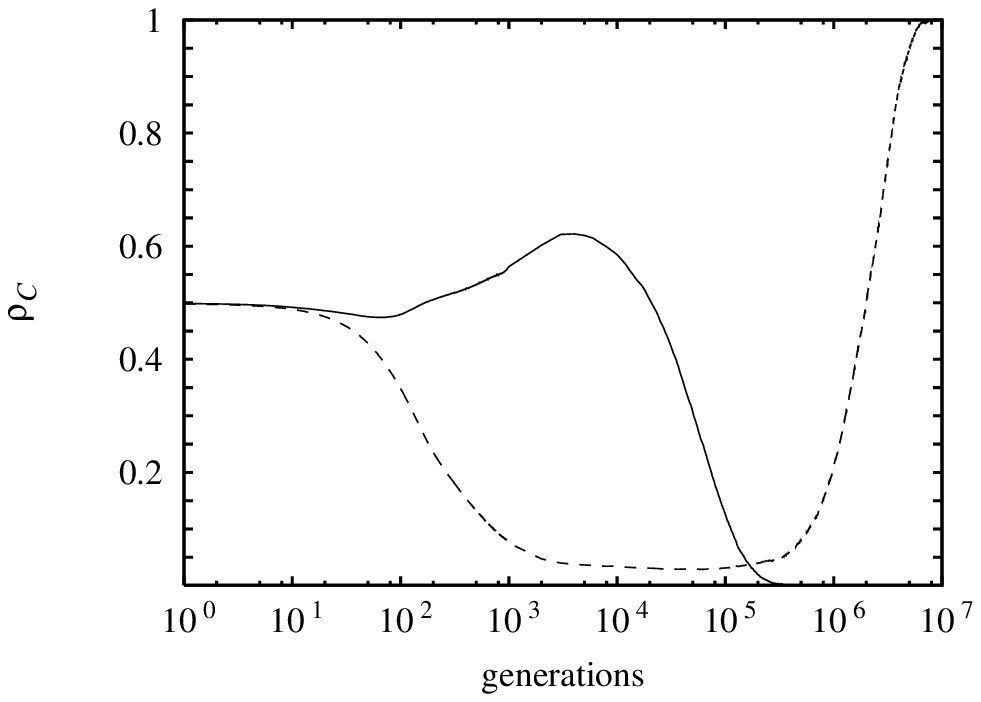}}
\caption{Exemplary time evolutions of $\rho_c$ for the three considered social dilemmas: (top) Prisoner's dilemma game ($T=1.2$, $S=0$, $K=2$); (middle) Snowdrift game ($T=1.8$, $S=0.8$, $K=2$); (bottom) Stag-hunt game ($T=0.9$, $S=-0.33$, $K=2$). In all panels solid and dashed lines depict results obtained via the coevolutionary promotion of cooperators (rule A) and defectors (rule B), respectively.}
\label{fig1}
\end{figure}

We start by presenting time evolutions of $\rho_C$ separately for the three considered social dilemmas in Fig.~\ref{fig1}, whereby payoff values are always set such that in the absence of coevolution cooperators could not survive. Clearly thus, the introduction of coevolutionary rules can promote cooperation, albeit the details depend somewhat on the type of the governing social dilemma as well as the applied rule (type A or B). In particular, while the coevolutionary promotion of defectors, introduced above as rule B, always sustains at least some fraction of cooperators (dashed lines), the coevolutionary promotion of cooperators (rule A) is ineffective in achieving the same goal (solid lines) as it fails to keep $\rho_C > 0$ for the same payoff values irrespective of which game applies. Nevertheless, it should be noted that the rather striking difference between rules A and B appearing by the stag-hunt game is mainly a consequence of the narrow region of a mixed phase at high noise levels rather then the difference in ability to promote cooperation, as we will show below.
Still, presented results convey persuasively that coevolutionary promoting defectors is more beneficial for cooperation than coevolutionary promoting cooperators. Importantly, this is true for the final outcome of all games, while at intermediate times it seems that the explicitly promoted strategy will actually fare better. Note that in all panels solid lines record a noticeable increase of $\rho_C$ at intermediate times, which is a direct consequence of the explicit promotion of cooperative behavior via the coevolutionary rule A. Likewise, all dashed lines depict similarly well-expressed drops of $\rho_C$, which is again a direct consequence of the applied rule B explicitly favoring defectors. Yet rather mysteriously, the tide always shifts in favor of the strategy that is not affected by the coevolution. It remains of interest to elaborate on the cooperation-promoting abilities of the two coevolutionary rules, and to explain why rule B is more successful in the long run.

\begin{figure}
\resizebox{0.95\columnwidth}{!}{\includegraphics{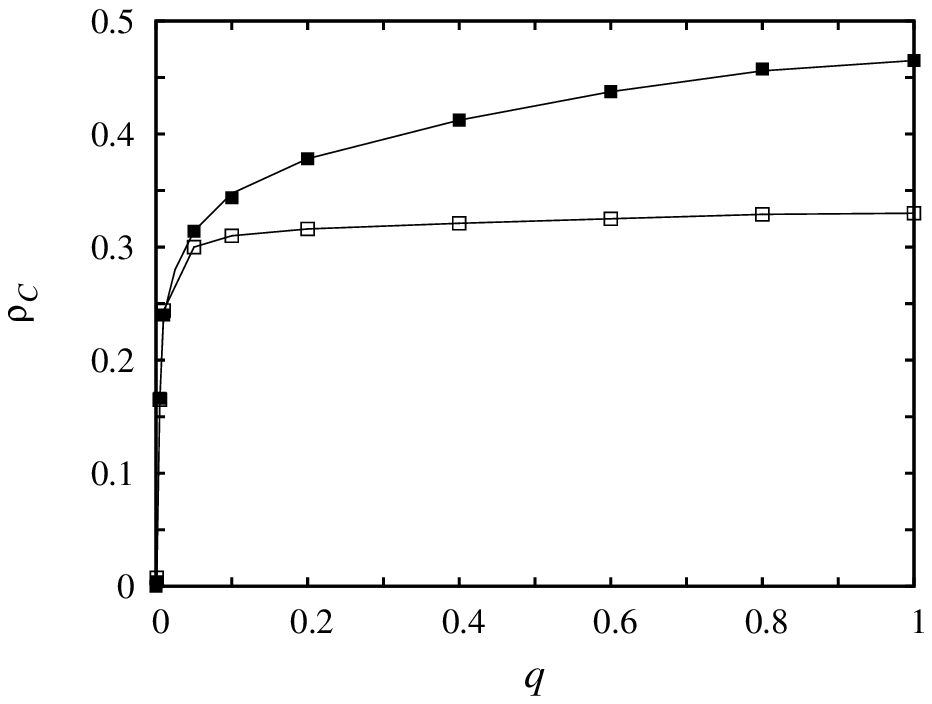}}
\caption{Cooperation level $\rho_C$ in dependence on the time separation between strategy and teaching activity updating $q$ for the prisoner's dilemma game ($T = 1.05$, $S=0$, $K=0.1$). Open and closed squares show stationary $\rho_c$ obtained via the coevolutionary promotion of cooperators (rule A) and defectors (rule B), respectively. Presented results are not relevantly affected by differences between the considered social dilemmas.}
\label{figad}
\end{figure}

But before examining the outlined facilitative effect of coevolution on cooperation in Fig.~\ref{fig1} more precisely, we first test the results against the separation of time scales \cite{sanchez_prl}, presently characterizing the evolution of strategies and teaching activity. By the model described in Section 2 the two time scales were treated as identical since, depending on the strategy and the applied coevolutionary rule (A or B), every successful reproduction was followed by an increase in the player's teaching activity. This model can be easily generalized via a parameter $q$ that determines the probability of increasing the teaching activity after each successful strategy pass. Evidently, $q=1$ recovers the originally proposed coevolutionary model while decreasing $q$ result in increasingly separated time scales. At $q=0$ the model becomes equivalent to the spatial model without coevolution, hence yielding $\rho_C = 0$ by high-enough $T$, as demonstrated in Fig.~\ref{figad}. On the other hand, an increase in $q$, resulting in a moderately fast yet effective coevolution, quickly becomes beneficial for cooperation since it enables the emergence of relevant heterogeneities among the teaching activities of players. It can also be inferred that the separation of time scales is somewhat more important by the coevolutionary promotion of defectors (closed squares in Fig.~\ref{figad}), which indicates that the slower evolution of $w_x$ acts against the awarding process, and that it is thus optimal to keep the coevolutionary process paced similarly fast as the main evolution of strategies. This conclusion is fully supported by the results presented in Fig.~\ref{figad}, since $\rho_C$ increases rather steadily with increasing values of $q$. However, since for all $q > 0.4$ the stationary values of $\rho_C$ increase no more than 10 \% across the remaining span of $q$, we will for simplicity continue to use $q=1$ in what follows, starting with a more in-depth examination of above results.

\begin{figure}
\resizebox{0.91\columnwidth}{!}{\includegraphics{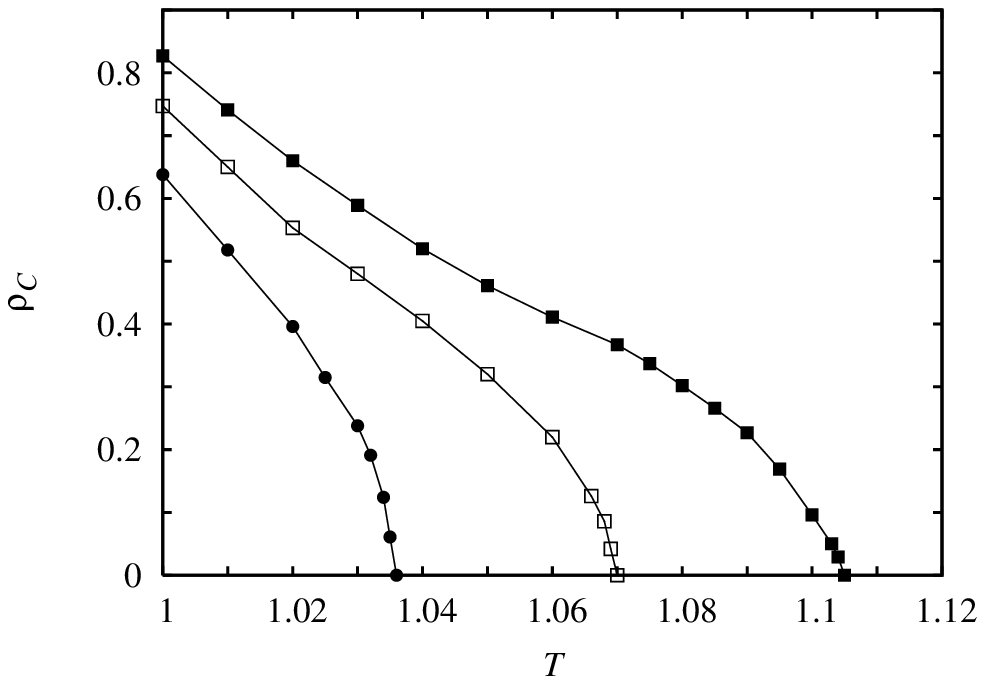}}
\resizebox{0.91\columnwidth}{!}{\includegraphics{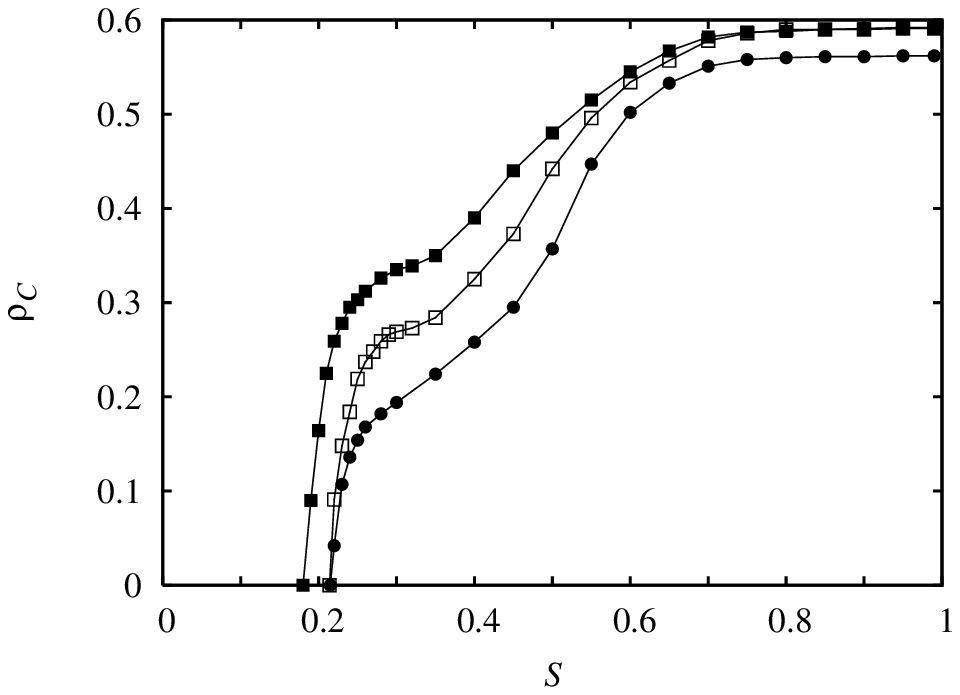}}
\resizebox{0.91\columnwidth}{!}{\includegraphics{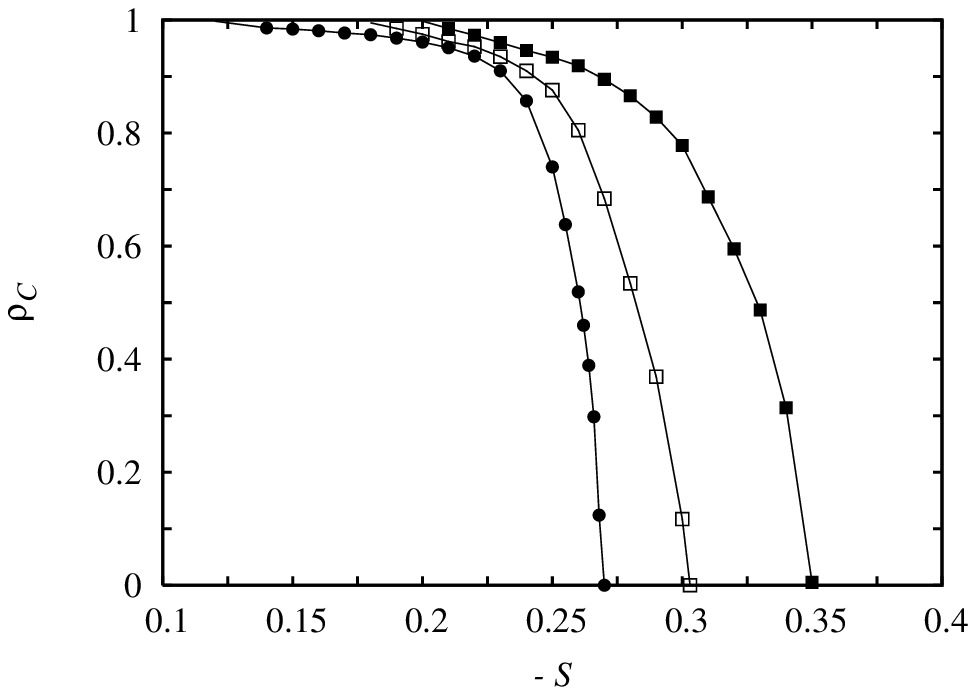}}
\caption{Promotion of cooperation via coevolution of teaching activity in the three considered social dilemmas: (top) Prisoner's dilemma game ($S=0$, $K=0.1$); (middle) Snowdrift game ($T=1.5$, $K=0.1$); (bottom) Stag-hunt game ($T=0.9$, $K=0.1$). In all panels closed circles depict results obtained with the classical version of the game (in the absence of coevolution), while open and closed squares show stationary $\rho_c$ obtained via the coevolutionary promotion of cooperators (rule A) and defectors (rule B), respectively.}
\label{fig2}
\end{figure}

In the following, to elaborate on the outlined cooperation-promoting effect in Fig.~\ref{fig1}, we present in Fig.~\ref{fig2} $\rho_C$ in dependence on the full relevant span of a given payoff for each considered social dilemma. Presented result fully support, first, that both considered versions of the coevolutionary rule do promote cooperation, and second, that rule B is more efficient in achieving this goal than rule A. In general, the impact of coevolution of teaching activity on cooperation is considerable, but it is also evident that the smallest impact can be detected when the snowdrift game applies.

\begin{figure}
\resizebox{0.95\columnwidth}{!}{\includegraphics{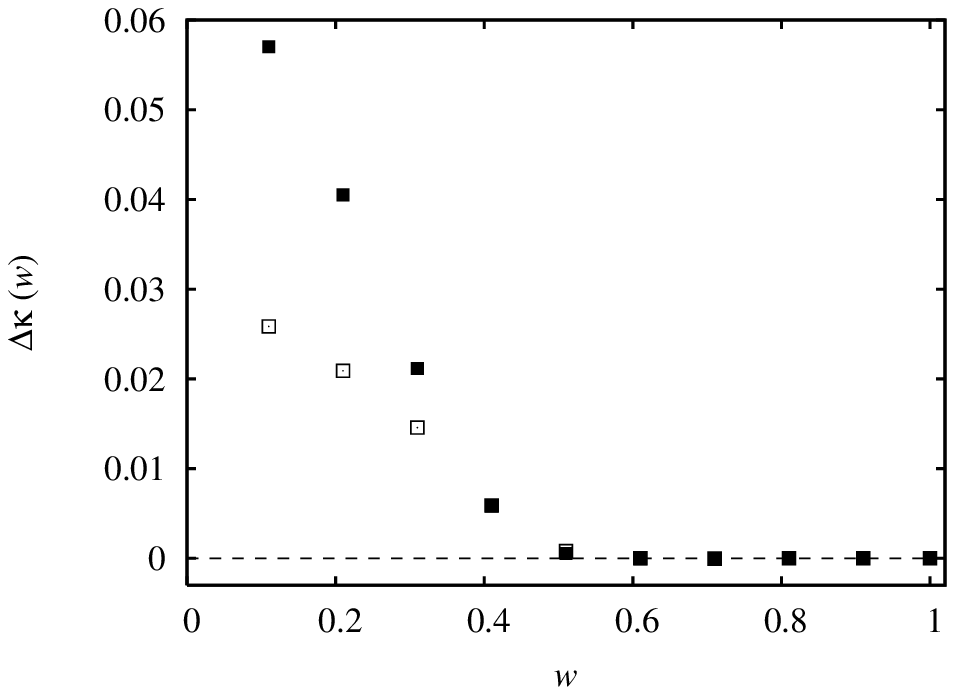}}
\caption{Differences in the distributions of teaching activity $\Delta \kappa(w)$ brought about by the two coevolutionary rules if $T=1.2$ and $S=0$ (prisoner's dilemma game). Closed and open squares depict results obtained by $K=0.1$ and $K=2$, respectively. Only active players having $w_x > w_m$ were considered. Presented results are not relevantly affected by differences between the considered social dilemmas.}
\label{fig3}
\end{figure}

To explain the shift in preference with respect to the explicitly promoted strategy exemplified at intermediate times in Fig.~\ref{fig1}, we examine the distributions of teaching activity $\kappa_{\rm A}(w)$ and $\kappa_{\rm B}(w)$ resulting from the application of coevolutionary rules A and B, respectively. More specifically, we focus on the difference $\Delta \kappa(w) = \kappa_{\rm B}(w) - \kappa_{\rm A}(w)$, which is presented in Fig.~\ref{fig3} for the prisoner's dilemma game at two different values of $K$. It can be observed that the B rule, explicitly promoting defectors, results in a larger fraction of players that are at least once affected by the coevolution ($w > w_m$) than the A rule. In fact, the difference is visible up to $w = 0.4$, thus indicating that the segregation of players into active (those having $w_x > w_m$) and virtually (or comparably) inactive (those having $w_x = w_m$) is noticeably stronger if the B rule is applied. This can be explained by acknowledging the fact that initially, {\it i.e.} in the random environment, the defectors are more successful, and thus rule B enables them to increase their teaching activity very efficiently. Furthermore, the strategy adoption process governed by Eq.~\ref{eq:prob} is more frequent when it is closer to the deterministic limit, and thus this is why the success of active players due to rule B is better pronounced at small values of the noise level $K$. Given the fact that substantial promotion of cooperation was in the past often associated with strongly heterogeneous states, for example in form of the host network \cite{santos05} or social diversity \cite{perc08}, it is reasonable to assume that the final segregation of players is responsible for the eventual shift in preference observed by the two coevolutionary rules in Fig.~\ref{fig1}, and also for the ultimately more potent promotion of cooperation via rule B, as demonstrated in Fig.~\ref{fig2}. In particular, we argue that the final distributions of $w_x$ are those having the decisive impact on the survival of the strategies, whereas during the coevolution itself (typically lasting around $100$ to $1000$ full Monte Carlo generations) the explicitly favored strategy simply enjoys a temporary uplift, which however, does not decisively determine its ultimate fate.

\begin{figure}
\resizebox{0.95\columnwidth}{!}{\includegraphics{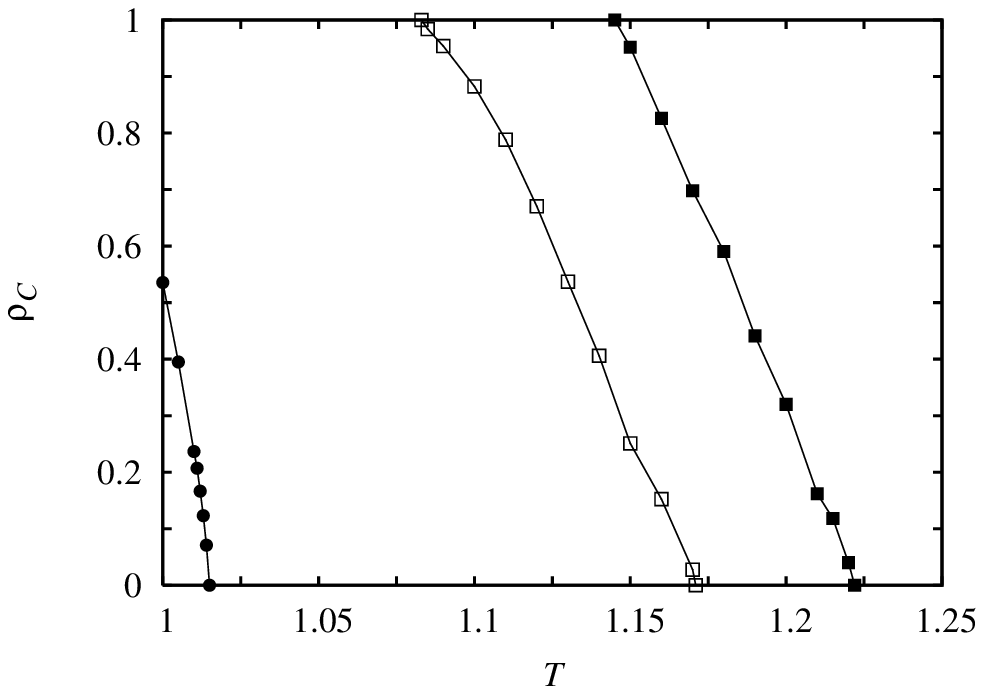}}
\caption{Promotion of cooperation via coevolution of teaching activity in the prisoner's dilemma game ($S=0$, $K=2$). Circles depict results obtained with the classical version of the game (in the absence of coevolution), while open and closed circles show stationary $\rho_c$ obtained via the coevolutionary promotion of cooperators (rule A) and defectors (rule B), respectively. It is instructive to compare these results with those presented in Fig.~\ref{fig2}(top).}
\label{fig4}
\end{figure}

By acknowledging the fact that differences between $\kappa_{\rm A}(w)$ and $\kappa_{\rm B}(w)$ are better expressed by lower intensities of selection $K$ (displayed in Fig.~\ref{fig3}), we can effectively support our reasoning by studying the impact of the two coevolutionary rules by different levels of uncertainty characterizing the strategy adoption process. Figure~\ref{fig4} features the same results as present in Fig.~\ref{fig2}(top), only that now $K=2$ instead of $K=0.1$ was used. According to our above arguments, the larger values of $\Delta \kappa(w)$, observed by $K=0.1$ in Fig.~\ref{fig3} (compared to $K=2$), should consequently result also in a larger difference between the cooperation-promoting abilities of rules A and B. Indeed, although the overall promotion of cooperation with respect to the classical version of the prisoner's dilemma game is better expressed by $K=2$ than $K=0.1$, the relative difference between rules A and B is clearly larger by $K=0.1$. Qualitatively identical results can be obtained also for other types of games, thus confirming that the decisive impact on the evolutionary success of the two strategies is issued, not by the explicit coevolutionary promotion of a given strategy, but by the final distribution of teaching activity resulting from the coevolutionary process. Since the coevolutionary promotion of defectors leads to a stronger segregation of players than the coevolutionary promotion of cooperators (see Fig.~\ref{fig3}), the cooperative behavior ultimately fares better via rule B. Although this fact might be temporarily masked by the explicit promotive nature of a given strategy due to the workings of the coevolutionary rule, as exemplified in Fig.~\ref{fig1}, eventually the active players seize full control over the game and reveal the true impact of coevolution.

\begin{figure}
\resizebox{0.95\columnwidth}{!}{\includegraphics{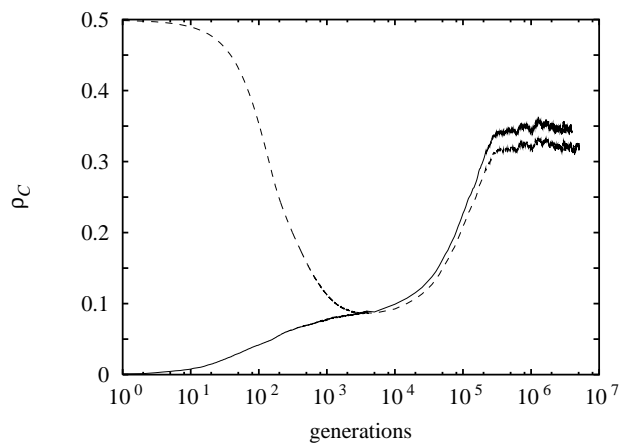}}
\caption{Temporal courses of $\rho_c$ for the whole population (dashed line) and for within the group of active players having $w_x > w_m$ (solid line). The prisoner's dilemma game [identical as in Fig. 1(top)] and the coevolutionary rule B (coevolutionary promotion of defectors) are applied. Clearly, the active players determine the final outcome of the game.}
\label{fig5}
\end{figure}

We argue that the increase of cooperation after a temporary setback period indicates that the promotive impact of the coevolutionary process is driven by a positive feedback mechanism. Notably, a short-term decline of cooperation has also been observed when the interaction graph of players was characterized by the scale-free topology (see Fig.~5 in Ref.\cite{santos_jeb06}). In the latter case a defector hub eventually becomes weak due to its predominantly defecting environment, which in turn impairs its ability to retain the defecting strategy. When the hub becomes occupied by a cooperator the overall cooperation level rises rapidly due to the efficient spreading enabled by the high degree of the prime spot of the network \cite{szabopr06,santos_prsb06}. As it was shown previously, players characterized with higher teaching activities play a similar role as hubs in a heterogeneous network, in particular, since they also have the ability to exploit the feedback mechanism postulated by a defecting neighborhood \cite{sz08}. To validate these arguments for the present model we measure the density of cooperators not just for the whole population but also for within the group of active players, as presented in Fig.\ref{fig5} when rule B applies. Owing to the fact that thus only defectors can be supported by the coevolution via increasing $w_x$, the cooperation density $\rho_C$ amongst active players starts at zero. As time passes, however, the active players (initially all defectors) may accidentally change their strategy and accordingly $\rho_C$ can start growing. The growth becomes more pronounced when active defectors fail to continue to effectively exploit their neighbors. Then namely they can no longer defend their active positions and start loosing frequently to cooperators. And since cooperators are much more effective in sustaining these prime (active) spots of the grid than defectors \cite{sz08pa}, cooperation can from thereon spread to inactive players as well, and thus uphold the cooperative behavior even if the temptation to defect is large. Consequently, the overall cooperation level starts growing as well. It is worth noting that this mechanism directly implies that the density of cooperators amongst active players is higher than the overall, which is in accordance with what has already been observed in evolutionary settings where a similar feedback mechanism took effect \cite{sz08,sz08pa}.

\section{Summary}\label{sec:sum}
We study the impact of simple coevolutionary rules affecting the teaching activity of players indulging either into the prisoner's dilemma, the snowdrift or the stag-hunt game. Irrespective of the details constituting the governing social dilemma, we demonstrate that the coevolution of teaching activity yields excessive benefits for the cooperators, substantially surpassing those that can be expect from spatiality alone. Rather surprisingly thereby, we show that the coevolutionary promotion of defectors is, in the long run, a more potent promoter of cooperation than the coevolutionary promotion of cooperators. While in the later case cooperative behavior initially does seem to fare better, in the long run the stronger segregation of players brought about by the coevolutionary promotion of defectors gives it the winning edge. Indeed, we show that the decisive impact on the evolutionary success of the two strategies is not issued, as one might intuitively expect, by the explicit coevolutionary promotion of a given strategy, but by the final distribution of teaching activity resulting from the applied coevolutionary process. Finally, we explain the cooperation-facilitating effect by showing that the active players have the ability to uplift the seemingly doomed cooperators and restore a socially viable cooperative state, whereby exploiting their celebrated role of higher influence that resulted from the coevolution of teaching activity at the very infancy of the game.

Presented results reiterate the importance of influential individuals in social dilemmas and strongly support the fact that appropriate conditions can emerge spontaneously via simple coevolutionary rules, thus fating the society to a predominantly cooperative state even if temptations to defect are high. Nevertheless, care should be exercised when deciding who to promote as seemingly correct decisions may backfire, and indeed, it seems that sometimes letting the bad seeds to grow is just what eventually yields the desired garden.

\begin{acknowledgement}
The authors acknowledge support from the Hungarian National Research Fund (grant K-73449) and the Slovenian Research Agency (grant Z1-9629). We have also benefited from discussions with Gy{\"o}rgy Szab{\'o}.
\end{acknowledgement}

\end{document}